\documentclass[aps,twocolumn,pra,superscriptaddress,amsmath,showpacs,tightenlines]{revtex4}
\usepackage{epsfig,graphicx,times}
\usepackage{amstext}
\usepackage{amsmath}            
\usepackage{amssymb}            
\usepackage{graphicx}           
\usepackage{latexsym}
\usepackage{bm}
\begin{document}
\title{
Superconducting qubits can be coupled and addressed as trapped
ions}

\author{ Yu-xi Liu}
\affiliation{CREST, Japan Science and Technology Agency (JST),
Kawaguchi, Saitama 332-0012, Japan} \affiliation{Frontier Research
System, The Institute of Physical and Chemical Research (RIKEN),
Wako-shi, Saitama 351-0198, Japan}
\author{L. F. Wei}
\affiliation{CREST, Japan Science and Technology Agency (JST),
Kawaguchi, Saitama 332-0012, Japan} \affiliation{Frontier Research
System, The Institute of Physical and Chemical Research (RIKEN),
Wako-shi, Saitama 351-0198, Japan}
\author{J. R. Johansson}
\affiliation{Frontier Research System, The Institute of Physical
and Chemical Research (RIKEN), Wako-shi, Saitama 351-0198,
Japan}
\author{J. S. Tsai}
\affiliation{CREST, Japan Science and Technology Agency (JST),
Kawaguchi, Saitama 332-0012, Japan} \affiliation{Frontier Research
System, The Institute of Physical and Chemical Research (RIKEN),
Wako-shi, Saitama 351-0198, Japan} \affiliation{NEC Fundamental
Research Laboratories, Tsukuba, Ibaraki 305-8051, Japan}
\author{Franco Nori}
\affiliation{CREST, Japan Science and Technology Agency (JST),
Kawaguchi, Saitama 332-0012, Japan} \affiliation{Frontier Research
System, The Institute of Physical and Chemical Research (RIKEN),
Wako-shi, Saitama 351-0198, Japan} \affiliation{Center for
Theoretical Physics, Physics Department, Center for the Study of
Complex Systems, The University of Michigan, Ann Arbor, Michigan
48109-1040, USA}
\date{\today}

\begin{abstract}
Exploiting the intrinsic {\it nonlinearity} of superconducting
Josephson junctions, we propose a scalable circuit with
superconducting qubits (SCQs) which is very similar to the
successful one now being used for trapped ions. The SCQs are
coupled to the ``vibrational" mode provided by a superconducting
LC circuit or its equivalent (e.g., a SQUID). Both single-qubit
rotations and qubit-LC-circuit couplings/decouplings can be
controlled by the {\it frequencies} of the time-dependent
magnetic fluxes. The circuit is scalable since the qubit-qubit
interactions, mediated by the LC circuit, can be selectively
performed, and the information transfer can be realized in a
controllable way.


\end{abstract}

\maketitle
\pagenumbering{arabic}

\section{Introduction}

Superconducting quantum circuits with Josephson junctions are
currently studied for their potential applications in quantum
information processing~\cite{phy-today}. Quantum coherent
oscillations and conditional gate operations have been
demonstrated using two coupled superconducting charge
qubits~\cite{pashkin1,pashkin2}. For a circuit with two coupled
flux qubits, spectroscopic measurements show that it acts as a
quantized four-level system~\cite{majer}. Further, entanglement
has been experimentally verified in coupled flux~\cite{izmalkov1}
and phase~\cite{xu,berkley,mcdermott} qubits.

A major challenge for superconducting qubits (SCQs) is how to
design an experimentally realizable circuit where the couplings
for different qubits can be selectively switched on and off, and
then scaled up to many qubits. Although two-qubit gates can be
generated (see, e.g., Ref.~\cite{strauch}) with always-on interbit
couplings, it is still very difficult to scale up experimental
circuits~\cite{pashkin1,pashkin2,majer,izmalkov1,xu,berkley,mcdermott}.
Theoretically, the circuits (e.g.,
Refs.~\cite{makhlin,you,you1,you2,liue,blais,wallraff,cleland,wei})
can be scaled up via a common data bus (DB). The DB modes are
virtually excited (e.g., Refs.~\cite{makhlin,you})  or excited
(e.g., Refs.~\cite{you1,you2,liue,blais,wallraff,cleland,wei}). In
the former case~\cite{makhlin,you}, the effective qubit couplings
can be switched on and off by changing the magnetic flux through
the circuit within nanoseconds, which is a challenge for current
experiments. In the later case~\cite{you1,you2,liue,
blais,wallraff,cleland}, the qubit and the DB are coupled or
decoupled when they have the same (resonant) or different
(non-resonant) frequencies, realized via a sudden non-adiabatic
change of either the qubit or the DB eigenfrequency. This
introduces additional noise.

The superconducting Josephson junction is a key building block of
superconducting quantum circuits.  Nonlinearity is its intrinsic
characteristic. This nonlinearity can be used to adjust the
inter-qubit couplings~\cite{plourde1,plourde2,plourde3,kim} by
changing the current bias of the coupler, and thus cancelling the
direct mutual inductance between the qubits. It can also be used
to realize the switchable coupling between two inductively coupled
superconducting flux qubits via a variable-frequency magnetic
flux~\cite{liu2006}. Also, recently, the level quantization of the
LC circuit has been experimentally demonstrated~\cite{LC,ntt}.

Combining the variable-frequency-controlled coupling
approach~\cite{liu2006} and experimental achievements of the
quantum LC circuit~\cite{LC,ntt}, we now study a different
approach to realize scalable SCQs via a common DB, which is either
a quantum LC circuit or its equivalent, modeled by a harmonic
oscillator~\cite{LC,ntt}. The equivalent LC circuits can be either
a cavity field (e.g., ~\cite{you1,you2,liue}) realized by, for
instance, a one-dimensional transmission line
resonator~\cite{wallraff} or a superconducting loop with Josephson
junctions (e.g., a dc-based SQUID). More significantly, all SCQs
can work at their optimal points when the data bus is a
superconducting loop with Josephson junctions (this is not the
case with standard LC DBs). In our this novel approach, the
individual properties (e.g., eigenfrequencies) of the DB and SCQs
are {\it always fixed}, but the SCQ-DB {\it couplings} can be
conveniently {\it controlled by changing the frequencies} of the
applied {\it time}-dependent magnetic fluxes (TDMFs). This is
promising, because it is often easier to produce fast and precise
frequency shifts of the radio-frequency control signal in
experiments, as opposed to changing the amplitude of the dc
signal.

We should point out that in our proposal the quantum LC circuit or
its equivalent has to be excited when the information is
transferred from one qubit to another; therefore it is an active
element, not a passive one which is just virtually excited. Our
proposal can be essentially reduced to the one used for trapped
ions~\cite{sasura}. The SCQs are coupled to the ``vibrational"
mode provided by a superconducting LC circuit or its equivalent
(e.g., a SQUID). Both single-qubit rotations and qubit-LC-circuit
couplings/decouplings can be controlled by the {\it frequencies}
of the time-dependent magnetic fluxes.  It means that SQCs can be
coupled and separately addressed similarly to trapped ions. This
similarity is significant because trapped ions~\cite{sasura} are
further ahead, along the quantum computing roadmap. It is
important to stress that our theoretical model can well explain
the blue/red sideband excitations which have been experimentally
implemented in superconducting qubit circuits~\cite{LC,semba}.

\begin{figure}
\vspace{-0.5cm}
\includegraphics[bb=12 440 650 760, width=10 cm, height=4.5cm, clip]{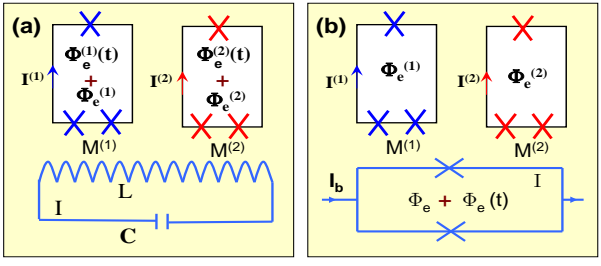}
\caption[]{(Color online) The $l$th flux qubit with three
junctions is coupled to an LC circuit in (a) or a dc-biased SQUID
with biased current $I_{b}$ in (b) by the mutual inductance
$M^{(l)}$ ($l=1, \,2$). (a) An externally applied magnetic flux
through the $l$th qubit loop includes a dc $\Phi_{e}^{(l)}$ term
and ac $\Phi_{e}^{(l)}(t)$ term controlling the coupling in (a).
The currents through the first qubit, second qubit, and LC circuit
in (a) [or SQUID loop in (b)] are $I^{(1)}$, $I^{(2)}$, and $I$
respectively. (b) However, when a dc-biased SQUID forms an
equivalent LC circuit, the SQUID-qubit couplings are controlled by
a TDMF, $\Phi_{e}(t)=A\cos(\omega_{c} t)$, through the SQUID loop.
The TDMF is added to the nonlinear qubit in (a) and to the
nonlinear SQUID loop in (b). The configuration in (b) is
significantly better, because both qubits can work at the optimal
point $f=1/2$. }\label{fig1}
\end{figure}

\section{Model} We study three-junction flux
qubits (e.g., ~\cite{orlando,yuxi}). As shown in Fig.~\ref{fig1},
we consider the simplest circuit, where two flux qubits are
coupled to a DB: either an LC circuit or a superconducting loop
with junctions (e.g., a dc-biased SQUID). Without loss of
generality and for simplicity, the DB is here assumed to be an LC
circuit with an inductance $L$ and a capacitance $C$. The mutual
inductance between the $l$th qubit and the LC circuit is $M^{(l)}$
($l=1,\,2$). The applied magnetic flux $\Phi^{(l)}$ through the
$l$th qubit loop in Fig.~\ref{fig1}(a) is assumed to include a
static (or dc) magnetic flux $\Phi_{\rm e}^{(l)}$, and also a TDMF
\begin{equation}
\Phi_{\rm e}^{(l)}(t)=A_{l}\cos(\omega^{(l)}_{c}
t),
\end{equation}
which controls the qubit-DB couplings.
Neglecting the mutual inductance between the two qubits, the
Hamiltonian can be written as
\begin{equation}\label{eq:1}
H=\sum_{l=1}^{2}H_{l}+\frac{Q^2}{2C}+
\frac{\phi^2}{2L}+\sum_{l=1}^{2} I M^{(l)} I^{(l)},
\end{equation}
with the current $I$ and magnetic flux $\phi=IL$ through the LC
circuit loop. Considering a three-junction qubit, the single-qubit
Hamiltonian $H_{l}$ in Eq.~(\ref{eq:1}) should be
\begin{equation}
H_{l}=\sum_{i=1}^{3}\frac{\Phi_{0}}{2\pi}\left[\frac{\Phi_{0}C^{(l)}_{{\rm
J}i}}{\pi}(\dot{\varphi}_{i}^{(l)})^2-I_{0i}^{(l)}\cos\varphi_{i}^{(l)}\right],
\end{equation}
after neglecting the qubit self-inductance and constant terms
$I_{0i}^{(l)}\Phi_{0}/2\pi\equiv E_{{\rm J}i}^{(l)}$. Each junction
in the $l$th qubit has a capacitance $C^{(l)}_{{\rm J}i}$, phase
drop $\varphi_{i}^{(l)}$, and supercurrent
$I_{i}^{(l)}=I_{0i}^{(l)}\sin\varphi_{i}^{(l)}$, with critical
current $I_{0i}^{(l)}$. The loop current of the $l$th qubit is
\begin{equation}\label{eq:3-1}
I^{(l)}=C_{l}\sum_{i=1}^{3}\frac{I^{(l)}_{0i}}{C^{(l)}_{{\rm J}
i}}\sin\varphi_{i}^{(l)},
\end{equation}
 where
\begin{equation}
\frac{1}{C_{l}}=\frac{1}{C^{(l)}_{{\rm J}1}}+\frac{1}{C^{(l)}_{{\rm
J}2}}+\frac{1}{C^{(l)}_{{\rm J}3}},
\end{equation}
 with the convention $C^{(l)}_{{\rm J}3}=\alpha_{l}
C^{(l)}_{{\rm J}1}=\alpha_{l} C^{(l)}_{{\rm J}2}$, and
$\alpha_{l}<1$. The LC circuit can be modeled by a harmonic
oscillator described by the creation operator
\begin{equation}
a^\dagger=\frac{1}{\sqrt{2\hbar \omega C}}(\omega C\phi-iQ)
\end{equation}
and its conjugate $a$, with frequency $\omega=1/\sqrt{LC}$.
Considering the TDMF, the phase constraint condition~\cite{orlando}
through the $l$th qubit loop becomes
\begin{equation}\label{eq:6}
\sum_{i=1}^{3}\varphi^{(l)}_{i}+2\pi\left[f+\frac{\Phi^{(l)}_{e}(t)}
{\Phi_{0}}\right] =0
\end{equation}
with the reduced bias flux
\begin{equation}
f=(\Phi^{(l)}_{e}-M^{(l)}I)/\Phi_{0}.
\end{equation}
Here, the bias $f$ includes the flux $M^{(l)} I$, produced by the
LC circuit. Thus, in the qubit basis, Eq.~(\ref{eq:1}) becomes
\begin{eqnarray}\label{eq:2}
H&=&\frac{\hbar}{2}\sum_{l=1}^2\omega_{q}^{(l)}\sigma^{(l)}_{z}+\hbar\omega
a{^\dagger}a+\sum_{l=1}^{2}H_{\rm int}^{(l)} \nonumber\\
&+&\sum_{l=1}^{2}(\lambda_{l}\sigma^{(l)}_{-}+{\rm
h.c.})\cos(\omega^{(l)}_{c} t)\\
&-&\sum_{l=1}^{2}(a^{\dagger}+a)(\Omega_{l}\sigma^{(l)}_{-}+{\rm
h.c.})\cos(\omega^{(l)}_{c} t)\,\nonumber
\end{eqnarray}
after neglecting the constant terms. Here the Pauli operators of
the $l$th qubit are defined by
$\sigma_{+}^{(l)}=|e_{l}\rangle\langle g_{l}|$,
$\sigma_{-}^{(l)}=|g_{l}\rangle\langle e_{l}|$, and
$\sigma_{z}^{(l)}=|e_{l}\rangle\langle e_{l}|-|g_{l}\rangle\langle
g_{l}|$. The computational basis states of the $l$th qubit are
defined~\cite{orlando,yuxi}, for $\Phi_{e}^{(l)}(t)=0$, by the two
lowest eigenstates, $|0\rangle_{l}=|g_{l}\rangle$, and
$|1\rangle_{l}=|e_{l}\rangle$, of $H_{l}$ with the two independent
variables
$\varphi^{(l)}_{p}=(\varphi^{(l)}_{1}+\varphi^{(l)}_{2})/2$ and
$\varphi^{(l)}_{m}=(\varphi^{(l)}_{1}-\varphi^{(l)}_{2})/2$.

The first two terms in Eq.~(\ref{eq:2}) denote the free
Hamiltonians of both qubits and the LC circuit; $\omega_{q}^{(l)}$
is the transition frequency of the $l$th qubit. The always-on
interaction Hamiltonian between the $l$th qubit and the DB in the
third term of Eq.~(\ref{eq:2}) is
\begin{equation}
H_{\rm int}^{(l)}=(a^\dagger+a)(G_{l}\,\sigma_{-}^{(l)}+{\rm h.c})
\end{equation}
with the coupling constant
\begin{equation}
G_{l}=M^{(l)}\sqrt{\frac{\hbar\omega}{2L}}\;\left\langle
e_{l}\left|I^{(l)}_{0}\right|g_{l}\right\rangle.
\end{equation}
The fourth term in Eq.~(\ref{eq:2}) represents the interaction
between the $l$th qubit and its TDMF with the interaction strength
\begin{equation}
\lambda_{l}=A_{l}\langle e_{l}|I^{(l)}_{3}|g_{l}\rangle.
\end{equation}
The fifth term of Eq.~(\ref{eq:2}) is the {\it controllable
nonlinear interaction} among the $l$th qubit, the DB, and the
TDMF,  with the coupling strength
\begin{equation}
\Omega_{l}=\frac{4\pi^2 A_{l}M^{(l)}C_{l}}{C^{(l)}_{{\rm
J}3}\Phi^2_{0}}\sqrt{\frac{\hbar\omega}{2L}}\left\langle
e_{l}\left|\;E^{(l)}_{{\rm
J}3}\cos\varphi^{(l)}_{3}\;\right|g_{l}\right\rangle.
\end{equation}
This nonlinear interaction term between the $l$th qubit, the DB,
and the TDMF originates from the expansion of the loop current
$I^{(l)}$ of the $l$th qubit in Eq.~(\ref{eq:3-1}) to first order
on the small reduced flux $\Phi_{e}^{(l)}(t)/\Phi_{0}$ via the
phase constrain condition in Eq.~(\ref{eq:6}). Above, the TDMF
$\Phi_{e}^{(l)}(t)$ equals zero when calculating the coupling
strengths $G_{l}$, $\lambda_{l}$, and $\Omega_{l}$. That is,
$I_{0}^{(l)}$ and $I^{(l)}_{3}$ are supercurrents through the loop
and the third junction respectively when $\Phi_{e}^{(l)}(t)=0$.

\section{Switchable interaction between qubit and data bus}

We find that the Hamiltonian~(\ref{eq:2}) can be reduced to the
one used in trapped ions~\cite{sasura} if the always-on
interaction terms $H^{(l)}_{\rm int}$ can be neglected. This
approximation is valid~\cite{mcdermott} during the TDMF
operations, in the large detuning regime between any qubit (e.g.,
$l$th qubit) and the DB
\begin{equation}
\Delta_{l}=\omega^{(l)}_{q}-\omega\gg |G_{l}|
\end{equation}
which can be achieved when the circuit is initially fabricated.
Thus, neglecting the always-on coupling $H_{\rm int}^{(l)}$
between the data bus and the qubits, the Hamiltonian (\ref{eq:2})
is reduced to
\begin{eqnarray}\label{eq:3}
H&=&\hbar\omega
a{^\dagger}a+\frac{\hbar}{2}\sum_{l=1}^2\omega_{q}^{(l)}\sigma^{(l)}_{z}\nonumber\\
&+&\sum_{l=1}^{2}(\lambda_{l}\sigma^{(l)}_{-}+{\rm
h.c.})\cos(\omega^{(l)}_{c} t)\\
&-&\sum_{l=1}^{2}(a^{\dagger}+a)(\Omega_{l}\sigma^{(l)}_{-}+{\rm
h.c.})\cos(\omega^{(l)}_{c} t)\,\nonumber
\end{eqnarray}
which now has the same form as the one used for quantum computing
with trapped ions, in the standard Lamb-Dicke limit (see, e.g.,
~\cite{sasura}).

Therefore, the essential difference between our Hamiltonian in
Eq.~(\ref{eq:2}) and the one used for
experiments~\cite{wallraff,LC,ntt} is that: (a) the nonlinear
coupling between the data bus, qubits, and the classical field in
Eq.~(\ref{eq:2}) is very important for the superconducting case.
Using this term, we can explain the sideband transitions in the
experiments~\cite{LC,ntt}; (b) the always-on coupling
$H^{(l)}_{\rm int}$ between the qubits and the data bus should be
negligibly small in our proposal. Our theoretical model is in
contrast with those in Refs.~\cite{LC,ntt} where: (a) there is no
nonlinear coupling between the data bus, qubits, and the classical
field; (b) the always-on Hamiltonian $H^{(l)}_{\rm int}$ could not
be neglected. That is, in Refs.~\cite{LC,ntt}, the Hamiltonian is
just the usual Jaynes-Cummings model which cannot be directly used
to explain the sideband excitations, especially for the
experimental results in Refs.~\cite{LC,semba}.

Analogous to the case of trapped ions, in our proposed devices,
three-types of dynamical evolutions ({\it carrier process; red
sideband excitation; and blue sideband excitation}) can be produced
by the TDMF using the {\it frequency-matching} (resonant) condition
and neglecting all fast oscillating terms. These three dynamical
evolutions can be described as follows.

(i) If $\omega^{(l)}_{c}=\omega^{(l)}_{q}$, the qubit and the DB
evolve independently in the large detuning condition. The external
flux $\Phi^{(l)}_{e}(t)$ is only used to separately address the
$l$th qubit rotations. These rotations are governed by the
Hamiltonian
\begin{equation}
H^{(l)}_{c}=\lambda_{l}\sigma^{(l)}_{-}+{\rm h.c.},
\end{equation}
in the interaction picture and using the rotating-wave
approximation (RWA) (also for the $H^{(l)}_{r}$ and $H^{(l)}_{b}$
shown below). This is the so-called {\it carrier process} in the
trapped ions approach.

(ii) If the frequencies satisfy the condition
$\omega^{(l)}_{c}=\omega^{(l)}_{q}-\omega$, then the
$\Phi^{(l)}_{e}(t)$ assists the $l$th qubit to couple resonantly
with the DB. This is the {\it red sideband} excitation, governed by
the Hamiltonian
\begin{equation}
H^{(l)}_{r}=\Omega_{l}\,a^{\dagger}\,\sigma^{(l)}_{-}+{\rm h.c.}.
\end{equation}

(iii) In the {\it blue sideband} excitation, the frequencies
satisfy the condition $\omega^{(l)}_{c}=\omega^{(l)}_{q}+\omega$,
with the Hamiltonian
\begin{equation}
H^{(l)}_{b}=\Omega_{l}a\sigma^{(l)}_{-}+{\rm h.c.}.
\end{equation}

Based on above discussions, it can be easily found that our
derived Hamiltonian in Eq.~(\ref{eq:2}), reduced to
Eq.~(\ref{eq:3}), can naturally {\it explain experimental results
on the sideband excitations}. For example, in Ref.~\cite{semba},
the qubit and the DB frequencies are $14$ GHz and $4.3$ GHz,
respectively; the frequency $\omega_{c}$ for the red or blue
sideband excitation is $9.7$ GHz or $18.32$ GHz. However, the
Jaynes-Cummings model cannot be used to explain these experiments.
The qubit-DB couplings/decouplings can be {\it controlled} by
appropriately selecting the $\omega^{(l)}_{c}$ of
$\Phi^{(l)}_{e}(t)$ to match/mismatch the above {\it frequency}
conditions of the sideband excitations.

\section{Single- and two-qubit gates}

For the $l$th qubit, the carrier process described by $H^{(l)}_{c}$
can be used to perform the single-qubit rotation
\begin{equation}
U^{(l)}_{c}(\beta_{l},
\phi_{l})=\exp\left[-i\frac{\beta_{l}}{2}\left(e^{-i\phi_{l}}\sigma^{(l)}_{-}
+e^{i\phi_{l}}\sigma^{(l)}_{+}\right)\right].
\end{equation}
Here, $\beta_{l}=|\lambda_{l}|\tau/\hbar$ depends on the Rabi
frequency $|\lambda_{l}|/\hbar$ and duration $\tau$; $\phi_{l}$ is
related to the phase of the TDMF applied to the $l$th qubit. For
example, the phases $\phi_{l}=0$ and $\phi_{l}=3\pi/2$ correspond
to the rotations $R^{(l)}_{x}(\beta_{l})$ and
$R^{(l)}_{y}(\beta_{l})$, about the $x$ and $y$ axis, respectively.
Thus, any single-qubit operation can be realized by a series of
$R^{(l)}_{x}(\beta_{l})$ and $R^{(l)}_{y}(\beta_{l})$ rotations
with well-chosen different angles $\beta_{l}$.

Two-qubit gates can be obtained using two qubits interacting
sequentially with their DB as in Ref.~\cite{sasura}. There, the
controlled phase-flip and the controlled-NOT gates can be obtained
by three and five steps, respectively. Here, we only discuss the
difference between our proposal and the one used for trapped ions.
In our proposed circuit, the ratio $|G_{l}|/\Delta_{l}$ cannot be
infinitely small. Then, the uncontrollable qubit-DB interaction
$H_{\rm int}^{(l)}$ needs to be considered by the effective
Hamiltonian~\cite{vitali}
\begin{equation}
H^{(l)}_{e}=\hbar
\frac{|G_{l}|^2}{\Delta_{l}}\left[|e_{l}\rangle\langle
e_{l}|\,aa^{\dagger}-|g_{l}\rangle\langle
g_{l}|\,a^{\dagger}a\right]
\end{equation}
when the $l$th qubit is not addressed by the TDMF. After
including this effect, three pulses (successively applied to the
first, second, and first qubits) with durations $\tau_{1}$,
$\tau_{2}$, and $\tau_{3}$ (used to perform a controlled
phase-flip gate in Ref.~\cite{sasura}) will result in a
two-qubit gate $U_{\rm two}$. This can be expressed as
\begin{equation}\label{eq:4}
U_{\rm two}=\left(\begin{array}{cccc}1&0&0&0\\
0&\exp(-i\theta_{1})&0&0\\ 0&0& \exp(i\theta_{2}) &0\\
0&0&0 &-\exp(-i\theta_{3})
\end{array}\right)
\end{equation}
in the two-qubit basis $\{|g_{1}\rangle|g_{2}\rangle,\,
|g_{1}\rangle|e_{2}\rangle,\,|e_{1}\rangle|g_{2}\rangle,\,
|e_{1}\rangle|e_{2}\rangle\}$ with the parameters
\begin{eqnarray}
\theta_{1}&=&\frac{2|G_{2}|^2}{\Delta_{2}}\tau_{1},\\
\theta_{2}&=&\frac{|G_{2}|^2}{\Delta_{2}}\tau_{1}+
\frac{|G_{1}|^2}{\Delta_1}\tau_{2},\\
\theta_{3}&=&\frac{3|G_{2}|^2}{\Delta_{2}}\tau_{1}
+\frac{|G_{1}|^2}{\Delta_1}\tau_{2}.
\end{eqnarray}
The two-qubit gate $U_{\rm two}$ in Eq.~(\ref{eq:4}) is just a
controlled phase-flip gate when the large detuning condition
$|G_{l}|/\Delta_{l}\sim 0$ is satisfied. Moreover, any quantum
operation can also be realized  by combining the two-qubit gate
$U_{\rm two}$ with other single-qubit operations.

\section{Entanglement and state transfer}

We now consider two different external fields satisfying
frequency-matching conditions, e.g., for the red sideband
excitation, which are simultaneously applied to the two qubits in
Fig.~\ref{fig1}. Then, in the interaction picture and the RWA, the
interaction Hamiltonian in Eq.~(\ref{eq:2}), between the LC circuit
and the two qubits, becomes
\begin{equation}
H_{1}=\sum_{l=1}^{2} (\Omega a^{\dagger}\sigma_{-}+{\rm h.c.}).
\end{equation}
For simplicity,  the coupling strengths between the LC circuit and
different qubits are now assumed to be identical, e.g.,
$\Omega_{1}=\Omega_{2}=|\Omega|e^{-i\theta}$. If the LC circuit is
initially prepared in the first excited state $|1\rangle$, then the
wave-function $|\Psi(t)\rangle$ of the whole system can be written
as
\begin{eqnarray}
|\Psi(t)\rangle&=&-ie^{i\theta}\sin\left(\sqrt{2}\Omega
t\right)[|e_{1}\rangle|g_{2}\rangle
|0\rangle+|g_{1}\rangle|e_{2}\rangle
|0\rangle]\nonumber\\
&+&\cos\left(\sqrt{2}\Omega
t\right)|g_{1}\rangle|g_{2}\rangle|1\rangle.
\end{eqnarray}
When $\sqrt{2}\,\Omega\, t/\hbar=\pi/2$, then the LC circuit is in
the vacuum state $|0\rangle$  and a maximally entangled state
between two qubits can be generated as
\begin{equation}
|\Psi^{+}\rangle_{12}=\frac{1}{\sqrt{2}}[|e_{1}\rangle|g_{2}\rangle
+|g_{1}\rangle|e_{2}\rangle ].
\end{equation}

When adding one more qubit to Fig.~\ref{fig1}(a) or (b), an
unknown state
$|\psi\rangle=\beta_{1}|g_{1}\rangle+\beta_{2}|e_{1}\rangle$ in
the first qubit can be transferred to the third one using the
standard teleportation procedure: i) a maximally entangled state
$|\Psi^{+}\rangle_{23}=[|e_{2}\rangle|g_{3}\rangle
+|g_{2}\rangle|e_{3}\rangle ]/\sqrt{2}$ between the second and
third qubits is prepared using the same method outlined above;
ii) a CNOT gate $U^{(12)}_{\rm CNOT}$ is implemented for the
first and second qubits (here, the second one is the target);
iii) a Hadmard gate is implemented on the first one; iv) a
simultaneous measurement, which can now be done
experimentally~\cite{mcdermott}, is performed on the first and
the second qubits.  The four different measured results
\{$|e_{1},e_{2}\rangle$,  $|e_{1},g_{2}\rangle$,
$|g_{1},e_{2}\rangle$,  and $|g_{1},g_{2}\rangle$\} correspond
to four outputs \{$|\psi_{1}\rangle$, $|\psi_{2}\rangle$,
$|\psi_{3}\rangle$, and $|\psi_{3}\rangle$\} in the third qubit.
The unknown state in the first qubit can be transferred to the
third one when the measured result for the first and second
qubits is $|e_{1}, e_{2}\rangle$. However, appropriate gates
(i.e., $\sigma^{(3)}_{x}$, $\sigma^{(3)}_{z}$, and
$\sigma^{(3)}_{z}\sigma^{(3)}_{x}$) need to be performed on the
other three outputs mentioned above to transfer $|\psi\rangle$
to the third qubit.
\begin{figure}
\includegraphics[bb=40 190 570 580, width=8 cm, clip]{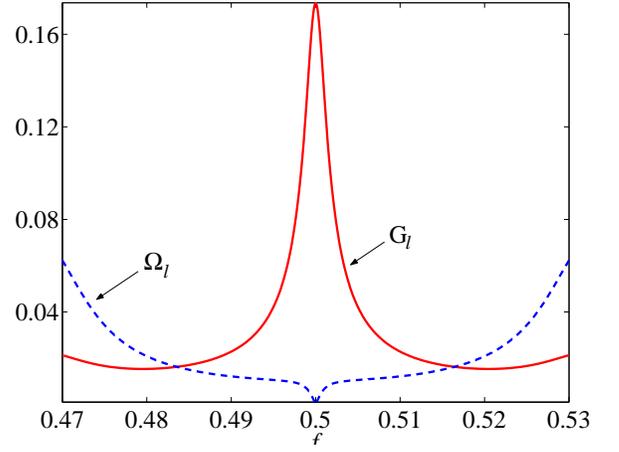}
\caption[]{(Color online) Plots of the $f$-dependent always-on
qubit-DB coupling strength $G_{l}$ (red curve) and the
TDMF-controlled qubit-DB coupling strength $\Omega_{l}$ (blue
curve), rescaled by $R_{l}=(2\pi/\Phi_{0})M^{(l)}E^{(l)}_{{\rm
J}3}\sqrt{\hbar\omega/2L}$.}\label{fig2}
\end{figure}

\section{Experimentally accessible parameters}

We now analyze the coupling constants related to the $l$th qubit:
(i) the always-on qubit-DB coupling strength
\begin{equation}
G_{l}\propto \langle e_{l}|I^{(l)}_{0}|g_{l}\rangle,
\end{equation}
and (ii) the TDMF-controlled qubit-DB coupling strength
\begin{equation}
\Omega_{l}\propto \langle e_{l}|\cos(2\varphi_{p}+2\pi
f)|g_{l}\rangle.
\end{equation}

At the degeneracy point $f=1/2$, the qubit potential is
symmetric~\cite{yuxi} and its ground and excited states have
opposite parities; however, $\cos(2\varphi_{p}+2\pi f)$ and the
qubit loop current $I^{(l)}_{0}$ have even and odd parities,
respectively. Therefore $\Omega_{l}=0$ but $G_{l}\neq 0$ when
$f=1/2$. Clearly, that $\Omega_{l}=0$ can be avoided by slightly
shifting $f$ away from the degeneracy point. The experiments on
sideband excitations, e.g., in Refs.~\cite{LC,semba}, were
performed with $f\neq 1/2$. Moreover, the controlled phase-flip
gate~\cite{sasura}, requiring a transition from the ground state
to the second excited state, also implies that the reduced
bias~\cite{yuxi} flux $f\neq 1/2$.

Figure~\ref{fig2} shows the $f$-dependent coupling strengths
$G_{l}$ and $\Omega_{l}$, rescaled by
\begin{equation}
R_{l}=(2\pi/\Phi_{0})M^{(l)}E^{(l)}_{{\rm
J}3}\sqrt{\hbar\omega/2L}.
\end{equation}
 As in Ref.~\cite{LC},  the Josephson
energies of the two bigger junctions of the $l$th qubit are
$E^{(l)}_{{\rm J}1}=E^{(l)}_{{\rm J}2}=225$ GHz and the ratio
between the small and big junction areas is $\alpha_{l}=0.76$. The
ratio between the Josephson energy $E_{J1}^{(l)}$ and the charge
energy $E_{c}^{(l)}$ of the $l$th qubit is about $30.8$. Using the
qubit parameters listed above and also taking the amplitude
$A_{l}=\Phi_{0}/30$ of the TDMF applied to the $l$th qubit,
Fig.~\ref{fig2} shows that $G_{l}$ and $\Omega_{l}$ are comparable
when $f$ is away from $1/2$; e.g., $G_{l}\approx 0.0579 \,R_{l}$
and $\Omega_{l}\approx 0.0224\, R_{l}$ when $f=0.49$. The strength
$\Omega_{l}$ can also be larger than the strength $G_{l}$ in the
range, e.g., $0.47\lesssim f\lesssim 0.477$.

If the capacitance and inductance of the LC circuit is
taken~\cite{LC} as $12$ pF and $250$ pH, then the frequency
$\omega$ of the LC circuit is about $2.9$ GHz. When the mutual
inductance $M^{(l)}$ between the $l$th qubit and the LC circuit is
taken as $20$ pH, then $G_{l}\approx 37.6$ MHz and
$\Omega_{l}\approx 14.6$ MHz when $f=0.49$. The $l$th qubit
frequency computed is about $18$ GHz when $f=0.49$. Therefore, the
detuning between the $l$th qubit and the LC circuit is
$\Delta_{l}\approx 15.1$ GHz, and the ratio
$G_{l}/\Delta_{l}\approx 0.0015$. Indeed, the {\it always-on
coupling $G_{l}$ is negligibly small} when the $l$th qubit works
at $f=0.49$ for measuring the sideband excitations. The phase
corrections $\theta_{i}$ in Eq.~(\ref{eq:4}) should be very small
with short operation times for those qubits when no
$\Phi^{(l)}_{e}(t)$ is applied.

For the LC circuit, if its capacitance $C$ and inductance $L$
are assumed as $\sim 1$ pF and $\sim 10$ nH, respectively, then
the LC circuit plasma frequency can be $\sim 1.6$ GHz. The
linear dimension for the LC circuit can be $\lesssim 1$ cm. The
estimated distance for a negligible mutual inductance between
two nearest qubits is $\sim 200 \,\mu$m, and thus one DB can
approximately interact with $\sim 40$ qubits. Of course, the
larger $L$ of the LC circuit could have a larger linear
dimension (allowing, e.g., $L\sim 100$ nH), and then more
qubits, here about $400$, could interact with the LC circuit. In
practice, the superpositions of the ground and excited states
for an LC circuit decay on a time scale given by $1/RC$, here
$R$ is the residual resistance of the circuit and its radiation
losses.

\section{Discussions and conclusions}

For flux qubits, the single-qubit states can be measured by using,
e.g., either a tank circuit weakly coupled to the
qubit~\cite{izmalkov1} or a dc-SQUID~\cite{bertet}. If only a
single-qubit measurement can be done at a time~\cite{liu} or
simultaneous measurements can be done (e.g., as for phase qubits
~\cite{steffen}), then any unknown quantum state can be
reconstructed~\cite{liu,steffen} and the information of the qubits
can be read out.

In our proposal, two crucial points are: (1) the qubit and the LC
circuit data bus should initially have a large detuning, such that
their always-on coupling is negligibly small when the
TDMF-assisted qubit-DB coupling is implemented; (2) the
nonlinearity of the JJs is essential to achieve our goal: the
nonlinear coupling between these {\it three}: the qubit, DB, and
TDMF. Based on these two requirements, the circuit can be modified
according to different experimental setups, e.g., the LC circuit
can be replaced by a superconducting loop with JJs (e.g., a
dc-biased SQUID as in Fig.~\ref{fig1}(b)). Three-junction flux
qubits can also be replaced by other qubits~\cite{phy-today},
e.g., one- or four-junction flux qubits, phase qubits, or
charge-flux qubits. Although the self-inductances of the qubits
are neglected here, our method is still valid for the qubits with
nonzero self-inductances~\cite{alec}.

Our numerical calculations show that the TDMF-controlled coupling
strength $\Omega_{l}$ is not large enough to realize very fast
two-qubit operations when the DB is a simple LC circuit. In
principle, this problem could be solved by using a superconducting
loop with Josephson junctions (e.g., dc-SQUID in
Ref.~\cite{plourde1}) as a data bus instead of a simple LC
circuit. Thus, the TDMF can be applied to the DB loop and the
qubit can work at the optimal point; the DB-qubit always-on
coupling can be minimized to zero; the TDMF-controlled
coupling-strength can be large enough to realize fast two-qubit
operations. A more detailed study on this issue will be presented
elsewhere.

In conclusions, using the nonlinearity of the superconducting JJs,
we theoretically explain the sideband excitations for qubits
coupled to an  LC circuit and show how to scale these to many
qubits. In contrast to previous proposals
(e.g.,~\cite{makhlin,you,wallraff,cleland}), the properties (e.g.,
eigenfrequencies) of the qubits and the DB are {\it fixed} when
processing either the resonant coupling or the non-resonant
decoupling.  Also, the qubit-DB couplings/decouplings are
controlled {\it neither} by changing the magnetic flux through the
loop {\it nor} by changing the eigenfrequencies of the qubits (or
the DB). They are {\it only} controlled via the {\it frequency
shifts} of TDMFs, which is much easier to achieve experimentally.

\acknowledgments We thank J.Q. You, Y. Nakamura, Y.A. Pashkin, O.
Astafiev, and K. Harrabi for discussions. FN was supported in part
by the US National Security Agency (NSA), Army Research Office
(ARO), Laboratory of Physical Sciences (LPS), and the National
Science Foundation grant No. EIA-0130383.

\end{document}